\documentclass[conference]{IEEEtran}
\IEEEoverridecommandlockouts
\usepackage{cite}
\usepackage{amsmath,amssymb,amsfonts}
\usepackage{algorithmic}
\usepackage{graphicx}
\usepackage{textcomp}
\usepackage{xcolor}
\usepackage{url}
\usepackage{array}
\usepackage{booktabs}

\begin{document}

\title{Hard-Earned Lessons in Access Control at Scale: Enforcing Identity and Policy Across Trust Boundaries with Reverse Proxies and mTLS}

\author{\IEEEauthorblockN{Mitendra Kumar Mahto}
\IEEEauthorblockA{\textit{Software Engineering} \\
\textit{LinkedIn Corporation}\\
Email: mahto.mitendra@gmail.com}
\and
\IEEEauthorblockN{Sanjay Singh}
\IEEEauthorblockA{\textit{Software Engineering} \\
\textit{LinkedIn Corporation}\\
Email: gargwanshi.sanjay@gmail.com}
}

\maketitle

\begin{abstract}
In today's enterprise environment, traditional access methods such as Virtual Private Networks (VPNs) and application-specific Single Sign-On (SSO) often fall short when it comes to securely scaling access for a distributed and dynamic workforce. This paper presents our experience implementing a modern, Zero Trust-aligned architecture that leverages a reverse proxy integrated with Mutual TLS (mTLS) and centralized SSO, along with the key challenges we encountered and lessons learned during its deployment and scaling. This multidimensional solution involves both per-device and per-user authentication, centralized enforcement of security policies, and comprehensive observability, hence enabling organizations to deliver secure and seamless access to their internal applications.
\end{abstract}

\begin{IEEEkeywords}
Zero Trust, mTLS, reverse proxy, access control, enterprise security, SSO
\end{IEEEkeywords}

\section{Introduction}
Modern enterprises face mounting complexity in securing access to their internal applications. The rise of remote work, bring-your-own-device (BYOD) policies, and an explosion of microservices has pushed legacy access control mechanisms to their limits. VPNs, once sufficient for perimeter-based security and primarily designed for remote access, now represent a broad and often risky gateway into corporate networks. Employees on campus, often on the same trusted network, can bypass VPN altogether and access internal resources directly, which is usually without any meaningful access restrictions beyond the basic network segmentation.

On one hand SSO improves usability, but introduces management challenges and performance bottlenecks when deployed at scale. This paper builds on emerging Zero Trust practices, such as those seen in BeyondCorp~\cite{beyondcorp} and advocates for a practical enterprise-wide transition from piecemeal controls to a centralized model based on reverse proxies with mTLS and integrated SSO. This architecture enforces strong identity and context-based access at network boundaries and provides a unified access layer across applications with central security policies being enforced.

The key goals for this paper are to:
\begin{itemize}
\item Highlight the shortcomings of current access mechanisms when confronted with the complexity of thousands of users, diverse tools, and varied protocols.
\item Describe a robust, proxy-based approach that centralizes policy enforcement, enabling contextual access control based on both user and device identity.
\item Share the practical, hard-earned lessons from deploying and operating this architecture in a live production environment, spanning various teams and infrastructure layers.
\end{itemize}

\section{Current State: Limitations of Legacy Approaches}
Modern enterprises operate within increasingly complex and distributed environments, making traditional access controls inadequate for secure and scalable operations. Legacy models, often designed for perimeter-centric security, have significant inconsistencies and vulnerabilities in how they manage access based on user location and trust assumptions.

\begin{table}[h]
\centering
\caption{Evolution of Access Control Mechanisms}
\small
\begin{tabular}{|p{0.8cm}|p{1.8cm}|p{1.8cm}|p{1.6cm}|}
\hline
\textbf{Phase} & \textbf{Primary Mechanism} & \textbf{Identity Determination} & \textbf{Access Control} \\
\hline
Phase 1 & Network ACLs (VPN) & IP address only & Coarse-grained, IP-based \\
\hline
Phase 2 & Basic SSO & User identity & Moderate, app-specific \\
\hline
Phase 3 & Enhanced SSO + RBAC & User + Groups & Role-based, moderately fine-grained \\
\hline
\end{tabular}
\label{tab:evolution}
\end{table}

\subsection{Fragmented Trust Models}
A primary limitation of legacy access models is the fragmented trust model they impose. Typically, remote employees rely on Virtual Private Networks (VPNs) for secure connectivity. While VPNs provide encrypted tunnels, their access control mechanisms primarily operate at Layer 3 (network layer) using Access Control Lists (ACLs) based on IP. Hence, it can enforce restrictions based solely on IP addresses or subnets. This means that once a device is authenticated to the VPN, it often gains broad, unrestricted lateral access to internal services, creating a vast attack surface if the device is compromised.

In contrast, on-campus employees typically don't require VPN access. While critical environments such as production are protected through strict network segmentation, as discussed later in the paper, non-production use cases such as internal tools or staging environments often allow broader access based solely on presence within a trusted corporate network. Over time, the distinction between critical and non-critical resources can blur. For example, staging environments may contain data copied from production, and internal tools may handle sensitive workflows. In many cases, ad hoc or isolated security measures are applied to mitigate these risks, but such solutions are often not comprehensive or consistently enforced, leading to potential exposure of sensitive information.

\subsection{Coarse-Grained Network Segmentation}
To mitigate lateral movement, large enterprises commonly employ network segmentation, creating boundaries between environments such as development, production, corporate, finance, and third-party zones. However, these segments are typically enforced through static Layer 3 or Layer 4 network controls. These controls are again coarse-grained, operating at the network packet level without context about user identity, device posture, or specific application requirements.

Maintaining these static network policies at scale becomes a significant operational burden, leading to configurations that are either brittle or end up granting more access than necessary. Such rigidity fails to capture the dynamic nature of modern enterprise needs, where user roles change frequently, device trustworthiness varies, and access often needs to be highly specific. A modern alternative that addresses these shortcomings is identity-based and device-aware policy enforcement at the traffic edge layer, typically enabled via reverse proxies. This approach allows for dynamic, context-rich access decisions that are separate from rigid network topologies.

\subsection{Limitations of Standalone Access Technologies}

\subsubsection{VPNs: Secure Tunnel, Broad Access}
VPNs are effective at establishing encrypted communication between users and corporate networks, safeguarding data in transit. However, their reliance on coarse-grained Layer 3 ACLs for access control poses a significant challenge. Once a device connects to the network via VPN, it generally receives unrestricted lateral access to internal services. This broad access model lacks the enforcement of least privilege and drastically expands the potential attack surface in the event of a compromised device.

\subsubsection{SSO: Convenience with Complexity}
Single Sign-On (SSO) enhances user experience by offloading authentication to a centralized identity provider, allowing users to access multiple applications without repeated logins. However, in practice, each application typically integrates with the SSO provider independently, leading to fragmented implementations. This becomes especially problematic in large enterprises where internal tools are hosted across different domains, such as .fin.company.com for finance apps and .tools.company.com for internal tools. These domain patterns often emerge from legacy setups, multi-datacenter deployments, or organizational silos, and are rarely standardized. Since browser-based SSO relies on shared cookies and domain continuity, such fragmentation makes centralized SSO brittle, resulting in inconsistent session behavior, repeated logins, and operational overhead. It also complicates policy enforcement and centralized auditing across the application ecosystem.

\subsubsection{The VPN + SSO Combination: Fragmented and Fragile}
The combination of VPNs (for network connectivity via Layer 3 ACLs) and per-application SSO introduces a complex and brittle access model. This hybrid approach exacerbates issues such as bloated HTTP headers and cookies which lead to performance degradation and application errors. Hence, access control remains fragmented across different systems, with no central point for policy enforcement or a ``kill switch'' to revoke access uniformly across all applications in the event of a security incident. This piecemeal strategy ultimately fails to scale effectively with modern organization requirements for secure, agile, and centrally managed access.

\begin{figure}[h]
\centering
\includegraphics[width=0.9\columnwidth]{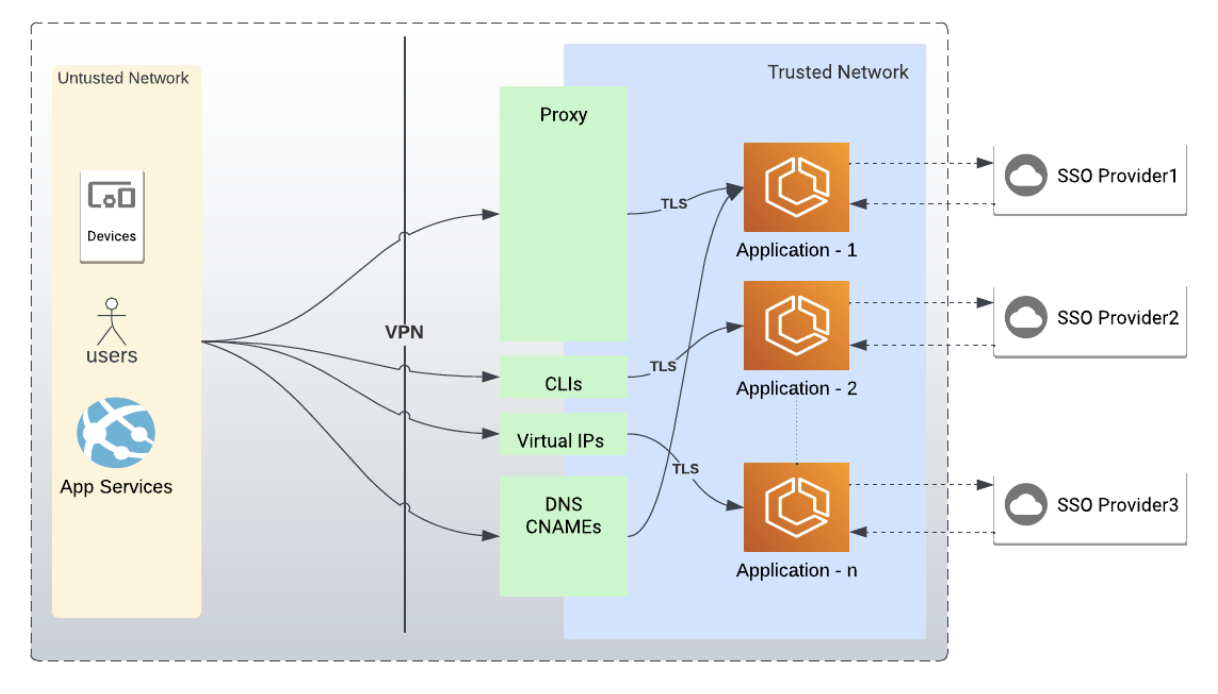}
\caption{Current State: Fragmented Access Control Landscape}
\label{fig:current-state}
\end{figure}

\section{Proposed State: The Zero Trust Model with Reverse Proxy, mTLS, and SSO}
The Zero Trust model represents a fundamental shift away from perimeter-based security by adopting the principle of ``never trust, always verify''~\cite{nist-zerotrust}. This philosophy replaces the traditional idea of a trusted network with a model of identity- and context-aware access enforcement. In this framework, every request for a resource, regardless of its origin, is treated as untrusted and must be verified in real time based on user identity, device trustworthiness, and strict policy compliance.

Our implemented architecture puts this philosophy into practice. It uses a reverse proxy integrated with Mutual TLS (mTLS) and centralized SSO. This proxy acts as a secure, unified gateway for all internal applications. Here's how it works:

\begin{enumerate}
\item It authenticates both the user and the device for every request using mTLS and SSO.
\item It evaluates a centralized set of access policies to determine if the request is permitted.
\item If access is granted, the proxy injects a single, lightweight identity header into the request. This header contains all the necessary validated information for the downstream application.
\end{enumerate}

By centralizing access decisions at the proxy, this architecture enables fine-grained access control, consolidated observability for auditing, and the ability to implement a rapid ``kill switch'' to revoke access across all applications in real time. This results in a more secure, scalable, and operationally simpler solution that still provides a seamless user experience.

We aim to equip infrastructure and security engineers with a practical framework and proven guidance for building Zero Trust-aligned access control systems that scale.

\begin{figure}[h]
\centering
\includegraphics[width=0.9\columnwidth]{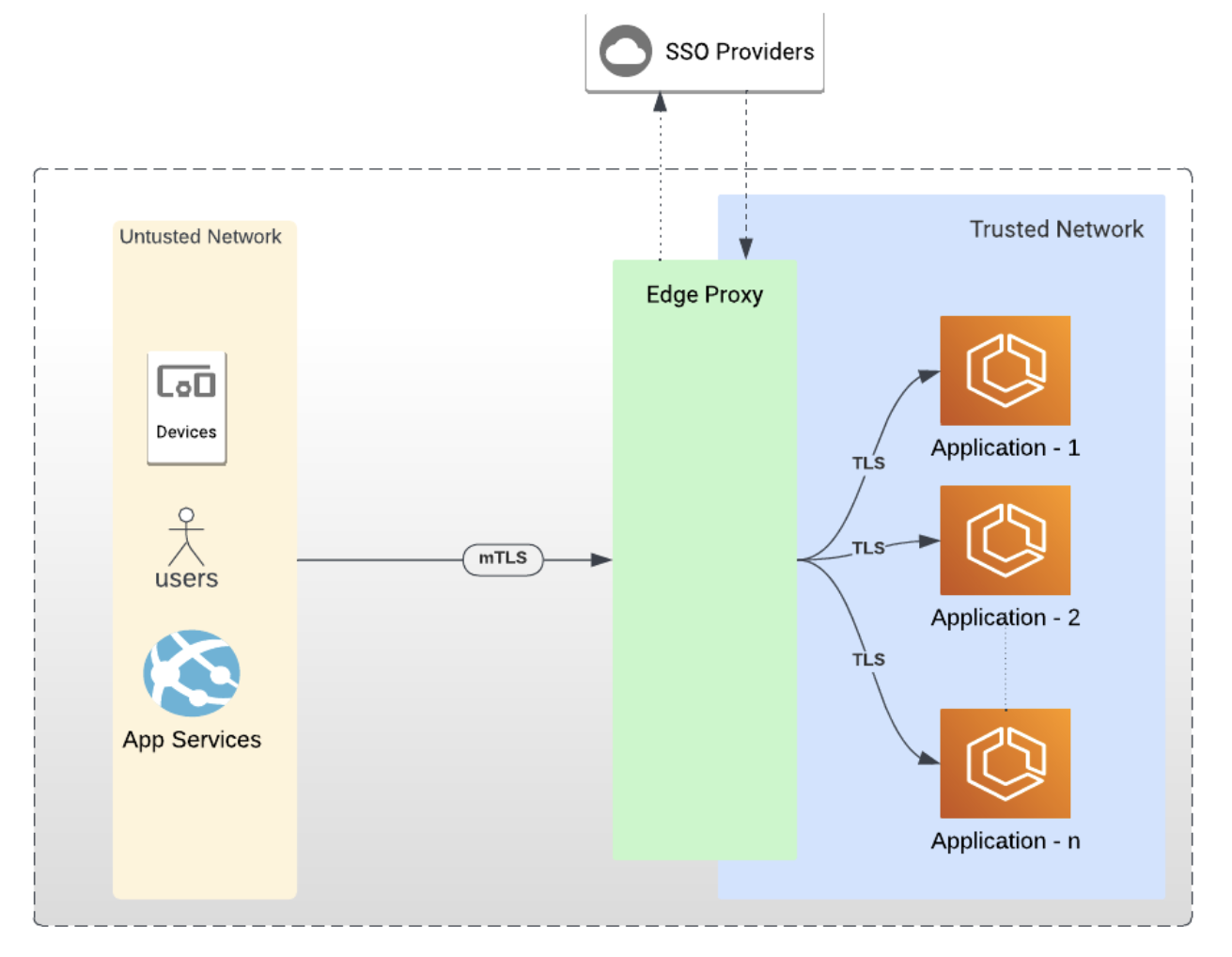}
\caption{Zero Trust Architecture with Reverse Proxy, mTLS, and SSO}
\label{fig:architecture}
\end{figure}

\section{System Architecture Overview}
At the heart of this architecture is a reverse proxy capable of performing both user and device authentication. This proxy acts as a critical enforcement point, performing both user and device authentication for every incoming request.

The proxy's operation is multi-fold:
\begin{itemize}
\item It terminates inbound HTTPS connections and leverages Mutual TLS (mTLS) to validate device certificates~\cite{mtls-guide}. This step ensures that only trusted, authenticated devices can initiate a connection.
\item Concurrently, it integrates with an identity provider (IdP) to perform robust user authentication using standard Single Sign-On (SSO) protocols such as OIDC or SAML. This verifies the identity of the individual attempting to gain access.
\item The proxy asynchronously retrieves policy enforcement data from an external control plane service. This allows it to apply broader organizational policies beyond basic user and device verification, such as detecting impossible login scenarios like simultaneous logins from geographically distant locations.
\item The proxy abstracts user authorization info in a secure token header as a standard JWT token signed by the proxy and passes it down to the applications. This decouples the application logic from SSO protocols and auth providers.
\end{itemize}

\subsection{Other Key Components}
To enable seamless integration across a diverse ecosystem, several supporting components were developed or extended:

\begin{itemize}
\item \textbf{Client-Side Libraries:} Libraries were implemented in Go, Rust, and Python to support mTLS and SSO flows across different platforms. For macOS specifically, integration with the native Keychain was required to authenticate device certificates securely.
\item \textbf{Command-Line Tool Enhancements:} Existing CLI tools were extended to support identity propagation by injecting token information into headers, enabling secure communication through the proxy.
\item \textbf{SSH-over-HTTP Integration:} SSH configurations were updated to enable tunneling over HTTP~\cite{ssh-http}. This allowed developers to access production environments via the reverse proxy, while enforcing mTLS and SSO authentication for all SSH sessions.
\item \textbf{Split-Horizon DNS Support:} Internal DNS infrastructure was enhanced to support Split-Horizon DNS, enabling context-aware routing through the appropriate proxy based on the client's security zone or environment.
\item \textbf{Browser Integration:} Browser behavior was extended to automatically retrieve the user's device certificate from secure storage (such as the macOS Keychain) and present it during mTLS handshakes, without prompting the user each time. This improved usability while maintaining strong authentication guarantees.
\item \textbf{Server-Side Token Libraries:} Server libraries were built in multiple languages to securely decode identity tokens and apply fine-grained authorization policies based on token claims and scopes.
\end{itemize}

\begin{figure}[h]
\centering
\includegraphics[width=0.9\columnwidth]{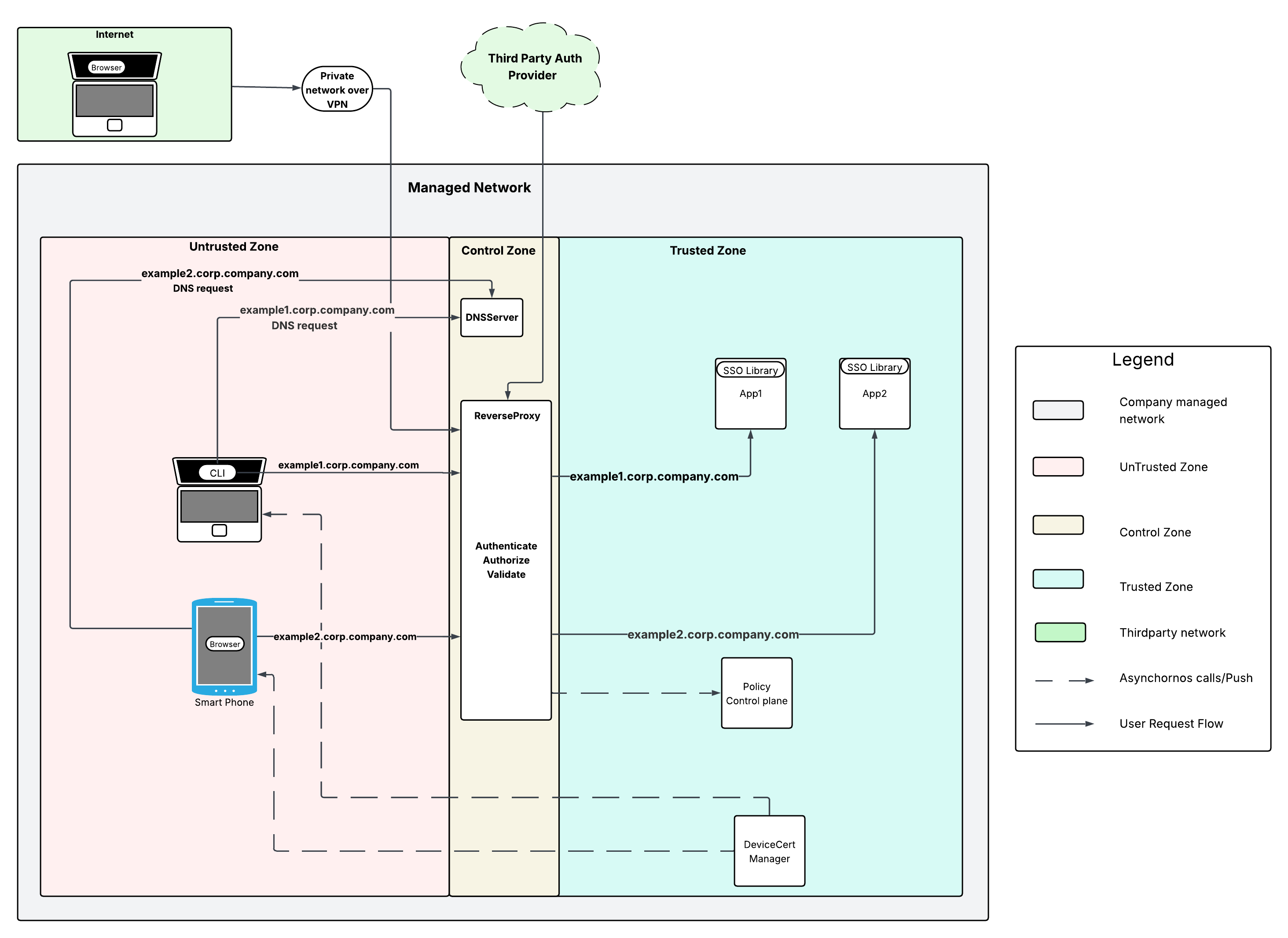}
\caption{Proxy Workflow and Component Integration}
\label{fig:proxy-workflow}
\end{figure}

\section{Detailed Request Flow}
This section outlines the step-by-step process a user's request follows when interacting with internal applications via the Zero Trust architecture.

\begin{enumerate}
\item \textbf{User Initiates Request:} A user types example1.corp.company.com into their browser.
\item \textbf{DNS Resolution:} The browser initiates a DNS query. Leveraging split-horizon DNS, the system returns the reverse proxy's IP address, directing traffic to our proxy.
\item \textbf{SSL Connection \& mTLS Handshake:} The browser attempts to establish an SSL/TLS connection with the reverse proxy. The proxy, in turn, initiates Mutual TLS (mTLS) by requesting a client certificate from the browser for device authentication.
\item \textbf{Device Certificate Presentation \& Validation:} The browser retrieves the client certificate from the macOS Keychain (or equivalent platform certificate store) and presents it. The proxy then validates this device certificate (checking against issuing CA, expiration, revocation status, etc.).
\item \textbf{Secure Connection Established:} Upon successful mTLS validation, a secure TCP connection is established between the browser and the proxy.
\item \textbf{HTTPS Request \& SSO Token Check:} The browser proceeds to send the HTTPS request over this established secure connection. The proxy intercepts the request and inspects it for a valid SSO token within the cookies.
\item \textbf{SSO Workflow Initiation:} If no valid SSO token is found, the proxy initiates the Single Sign-On (SSO) workflow with the configured authentication provider.
\item \textbf{User Authentication:} The user is redirected to the identity provider and completes the SSO workflow by entering their login credentials or by leveraging existing sessions.
\item \textbf{Dynamic Policy Enforcement:} After successful user and device authentication, the proxy applies additional security policies. It leverages dynamically updated policy enforcement data, retrieved asynchronously from an external control plane service. This allows for the application of broader organizational policies, such as impossible login, revoked device certs, etc, before granting final access to the internal application.
\item \textbf{Subsequent Authenticated Requests \& Downstream Forwarding:} Once all policy checks are passed and a valid SSO token is established, all subsequent requests from that browser will automatically include this token in the cookies. The proxy then extracts this SSO token from the cookie and sets it as a header for downstream applications, ensuring seamless and continually evaluated access.
\item \textbf{Logging and Observability:} Each access decision is logged with full context: certificate fingerprint, user ID, source IP, request path, and decision outcome. These logs are critical for security audits, forensic analysis, and compliance reporting.
\end{enumerate}

\section{Implementation Challenges}
This section outlines three major categories of challenges encountered while deploying a proxy-based Zero Trust architecture at scale: operational, architectural, and policy-related. Addressing these challenges holistically is essential to ensure successful and sustainable rollout. A particularly relevant example involves differentiating access between full-time employees (FTEs) and contractors. Contractors often use unmanaged devices, require limited access durations, or operate with reduced trust posture compared to full-time staff. Designing a policy framework that accounts for these nuances is critical. This is where device metadata and identity-based policy enforcement which are enabled through mTLS and SSO integration at the proxy to become indispensable.

\subsection{Operational Challenges: Certificate Lifecycle and Device Integration}
One of the foremost tasks is managing certificates at scale. Provisioning, rotating, and revoking certificates across thousands of devices requires a robust Public Key Infrastructure (PKI) tightly integrated with device management systems. Automation becomes key to reduce operational overhead and avoid security lapses.

Additionally, securing private key material on user devices is critical. Leveraging platform-native capabilities such as Secure Enclave on macOS or TPMs on Windows which ensures private keys are stored in hardware-isolated environments, following best practices for secure key handling~\cite{owasp-mtls}.

\subsection{Proxy Integration Challenges}
At the architectural level, integrating the reverse proxy into existing infrastructure poses its own complexities. Internal tooling and legacy applications may not support mTLS or header-based identity propagation, requiring custom integrations or shims. Client certificate handling is not equally mature across all programming languages, leading to inconsistencies and increased development effort. The proxy must also handle a variety of protocols and traffic types, sometimes necessitating advanced routing logic or fallback modes for edge cases.

\subsection{Policy Modeling and Backward Compatibility}
Designing access control policies that are expressive enough to handle modern identity-based models while remaining compatible with legacy systems is a non-trivial task. Policies must account for hybrid scenarios where tools lack full mTLS or SSO support, or where device trust cannot be independently established. Moreover, teams often interpret ``least privilege'' differently, so striking the right balance between strictness and usability requires iterative tuning and stakeholder buy-in.

\section{Lessons Learned}
Implementing a Zero Trust-aligned architecture with a reverse proxy, mTLS, and centralized SSO brought several critical insights. Our experiences highlight the importance of robust observability, differentiated policies, and resilient infrastructure.

\subsection{Careful Planning is Essential to Manage Co-existence Challenges}
Legacy, decentralized SSO often leads to excessive cookie and header sizes, causing HTTP 431 errors and performance degradation. While centralizing identity verification at the proxy solves these issues by removing the need to pass large tokens and simplifies integration with older systems, organizations must prepare for unexpected surprises during the transition. Apps that previously managed their own SSO and header controls might behave unpredictably when both old, decentralized methods and the new, centralized proxy-based controls are active. Proactive and thorough planning is crucial to navigate these complex co-existence scenarios and ensures a smooth migration.

\textbf{Recommendations:} Plan transitions meticulously. Audit applications for legacy header behavior, and coordinate closely with app teams to avoid conflicts between old and new SSO mechanisms. Use feature flags or phased rollout strategies to manage risk.

\subsection{Expect and Plan for Tooling and Platform Fragmentation}
Organizations often operate in heterogeneous environments, where various command-line tools, web UIs, and backend services interact differently with identity and certificate infrastructure. Client behaviors can vary widely based on language version (e.g., Python 2 vs 3) or operating system (e.g., macOS vs. Linux vs. Windows), as well as their support for certificate stores like macOS Keychain. For instance, macOS's strict Keychain enforcement can interfere with mTLS flows, often requiring device-specific workarounds. Accounting for this fragmentation is crucial for seamless deployment and user experience.

\textbf{Recommendations:} Provide platform-aware guidance and wrapper scripts. Proactively test enforcement across common stacks. Avoid assuming uniform behavior across developer environments.

\subsection{Lack of Observability Can Cripple Enforcement}
To successfully implement dynamic access decisions and enforce a deny-by-default posture, comprehensive and real-time observability is critical. This includes capturing and querying identity context, device trust signals, and access logs. This deep visibility allows security teams to confidently refine policy behavior, ensuring that access is granted exactly as intended, preventing both over-provisioning and unnecessary denials. Without this precise visibility and control, security teams can't be sure policies are working as intended, and users may face frustrating experiences due to unexpected access issues.

\textbf{Recommendations:} Add structured logging, user-facing audit trails, and rich error messages. Invest in tooling that can trace decisions across proxies and identity providers. Make observability a prerequisite, not an afterthought.

\subsection{The Proxy Must Be Treated as Tier-0 Infrastructure}
The proxy serves as the single enforcement point for access decisions across the organization, making it a critical dependency. Any disruption, whether caused by outages, misconfigurations, or failures in the identity provider, can lead to widespread access issues. Its central role significantly increases operational risk, especially when it controls entry to sensitive environments or core systems.

\textbf{Recommendations:} To mitigate these risks, the proxy path should be hardened with safeguards such as fallback authentication mechanisms, signed token caching, certificate monitoring, and high availability. Out-of-band access paths should be established for critical operations like SSH, logging, and incident response to maintain control during outages. The proxy and its observability stack must be treated as Tier-0 infrastructure, supported by safe CI/CD practices, automated configuration validation, and region-aware deployment strategies. Additionally, identity validation logic should be made testable and version-controlled, with support for staged rollouts, traffic shadowing, and rollback mechanisms to minimize disruption during updates.

\subsection{Support Protocol Diversity Without Forcing Rewrites}
Organizational needs for identity management can evolve rapidly, necessitating the ability to seamlessly switch between different SSO authentication protocols and identity providers. Our architecture's proxy provides a critical abstraction layer for identity that addresses this need. It ensures that internal applications, even those lacking native SSO or mTLS support, do not need to be aware of or adapt to changes in the specific authentication protocol (e.g., SAML, OIDC, Kerberos) or the underlying identity provider. Therefore, leveraging the proxy to handle all necessary protocol conversions and assertions is crucial for future-proofing an organization's identity strategy.

\textbf{Recommendations:} Normalize identity at the proxy. Accept multiple formats upstream and translate them into consistent headers or tokens downstream. Allow services to evolve independently of identity systems.

\section{Conclusion}
Transitioning to a reverse proxy model with mTLS and centralized SSO aligns with Zero Trust principles and offers a significant upgrade in access control capabilities. It enables granular, context-aware authentication decisions based on both user and device identity, simplifies application integration, and provides a centralized enforcement point for policy and observability. While not without its implementation challenges, the security, scalability, and operational benefits of this approach make it a compelling choice for modern enterprise environments.

\end{document}